\documentclass[%
 prl,
 amsmath,amssymb,
 reprint,%
]{revtex4-1}

\usepackage{graphicx}% Include figure files
\usepackage{dcolumn}% Align table columns on decimal point
\usepackage{bm}% bold math
\usepackage[utf8]{inputenc}
\usepackage[T1]{fontenc}
\usepackage{mathptmx}
\usepackage{etoolbox}
\usepackage{xcolor}

\makeatletter
\def\@email#1#2{%
 \endgroup
 \patchcmd{\titleblock@produce}
  {\frontmatter@RRAPformat}
  {\frontmatter@RRAPformat{\produce@RRAP{*#1\href{mailto:#2}{#2}}}\frontmatter@RRAPformat}
  {}{}
}%
\makeatother
\graphicspath{Figures/}

\begin{document}

\preprint{AIP/123-QED}

\title{Superconductivity in hyperdoped Ge by molecular beam epitaxy}
\author{Patrick J. Strohbeen}
\author{Aurelia M. Brook}
\affiliation{
Center for Quantum Information Physics, Department of Physics, New York University, New York, NY 10003 USA
}
\author{Wendy L. Sarney}
\affiliation{%
Army Research Directorate, DEVCOM Army Research Laboratory, Adelphi, MD 20783 USA
}%
\author{Javad Shabani}
\thanks{corresponding author: jshabani@nyu.edu}
\affiliation{Center for Quantum Information Physics, Department of Physics, New York University, New York, NY 10003 USA
}%

\date{\today}% It is always \today, today,

\begin{abstract}

Superconducting germanium films are an intriguing material for possible applications in fields such as cryogenic electronics and quantum bits. Recently, there has been great deal of progress in hyperdoping of Ga doped Ge using ion implantation. The thin film growths would be advantageous allowing homoepitaxy of doped and undoped Ge films opening possibilities for vertical Josephson junctions. Here, we present our studies on the growth of one layer of hyperdoped superconducting germanium thin film via molecular beam epitaxy. We observe a fragile superconducting phase which is extremely sensitive to processing conditions and can easily phase-segregate, forming a percolated network of pure gallium metal. By suppressing phase segregation through temperature control we find a superconducting phase that is unique and appears coherent to the underlying Ge substrate.

% However, the current state of the research has left many open questions as to the nature of this phase, how to control it, and how we can implement it in real device architectures.

\end{abstract}

\maketitle

High-quality thin films of covalent superconducting material is enticing for pushing towards up-scaling current state-of-the-art quantum devices. Proposals calling for highly scalable device platforms such as cryogenic CMOS circuitry \cite{patra2017cryocmos} or qubit architectures \cite{ShimTahan, zhao2020merge} have emerged as a new challenge in finding materials such as Si or Ge for devices with superior performance at cryogenic temperatures \cite{tahan2014bottomup}. Additionally, covalent superconductors offer a unique aspect in that high-quality fully homoepitaxial Josephson junctions can be created in the growth direction \cite{zhao2020merge, mcrae2021IIIVMerge}. Thus, drawing inspiration from the original superconducting qubits \cite{nakamura1999cpbox,vion2002quantronium} one can begin to consider qubit designs that make use of the out-of-plane degree of freedom such as the recently studied merged-element transmon, or ``Mergemon'' \cite{zhao2020merge}. Such designs have already shown through preliminary evaluations the impact of the superconductor-semiconductor interface on the observed information coherence times \cite{zhao2020merge,mcrae2021IIIVMerge}. It is projected that high quality epitaxial interfaces will solve this problem, putting forth a strong proposal for new material and device applications. Thus, further efforts must be made into developing highly compatible superconducting materials as qubit density requirements become more stringent. 

Superconductivity in group IV semiconductors has been presented in diamond \cite{ekimov2004superDiamond}, silicon \cite{bustarret2006superSilicon, sardashti2021supersi}, and germanium \cite{sardashti2021GaGeprm, fiedler2012GaGeSiO2capping, herrmannsdorfer2009superGaGe} previously, though these studies use high energy synthesis techniques (e.g. ion implantation) and many questions remain to be investigated. Particularly, when hyperdoping Ge with high energy Ga ion beams, the origin of superconductivity has remained controversial due to high energies driving possible phase segregation\cite{sardashti2021GaGeprm, skrotzki2011Gedoping, fiedler2012GaGeSiO2capping, herrmannsdorfer2009superGaGe, prucnal2019algaGe}. For example, typically a thick silicon oxide layer is first deposited to protect the surface during the implantation process. \cite{sardashti2021GaGeprm, sardashti2021supersi, fiedler2012GaGeSiO2capping}After implantation and subsequent flash annealing at high temperature to activate the Ga acceptors, superconductivity is attained. However, if the protective oxide layer is etched away, the superconducting state disappears.\cite{sardashti2021GaGeprm} This has resulted in the hypothesis that the Ga-metal is diffusing to the SiO$_{x}$/Ge interface and coalescing into a percolated network of metallic Ga, which is in itself an elemental BCS superconductor even in a confined state \cite{charnaya1998confinedga, charnaya2009gananoconfine}.

\begin{figure}
    \centering
    \includegraphics[width=\linewidth]{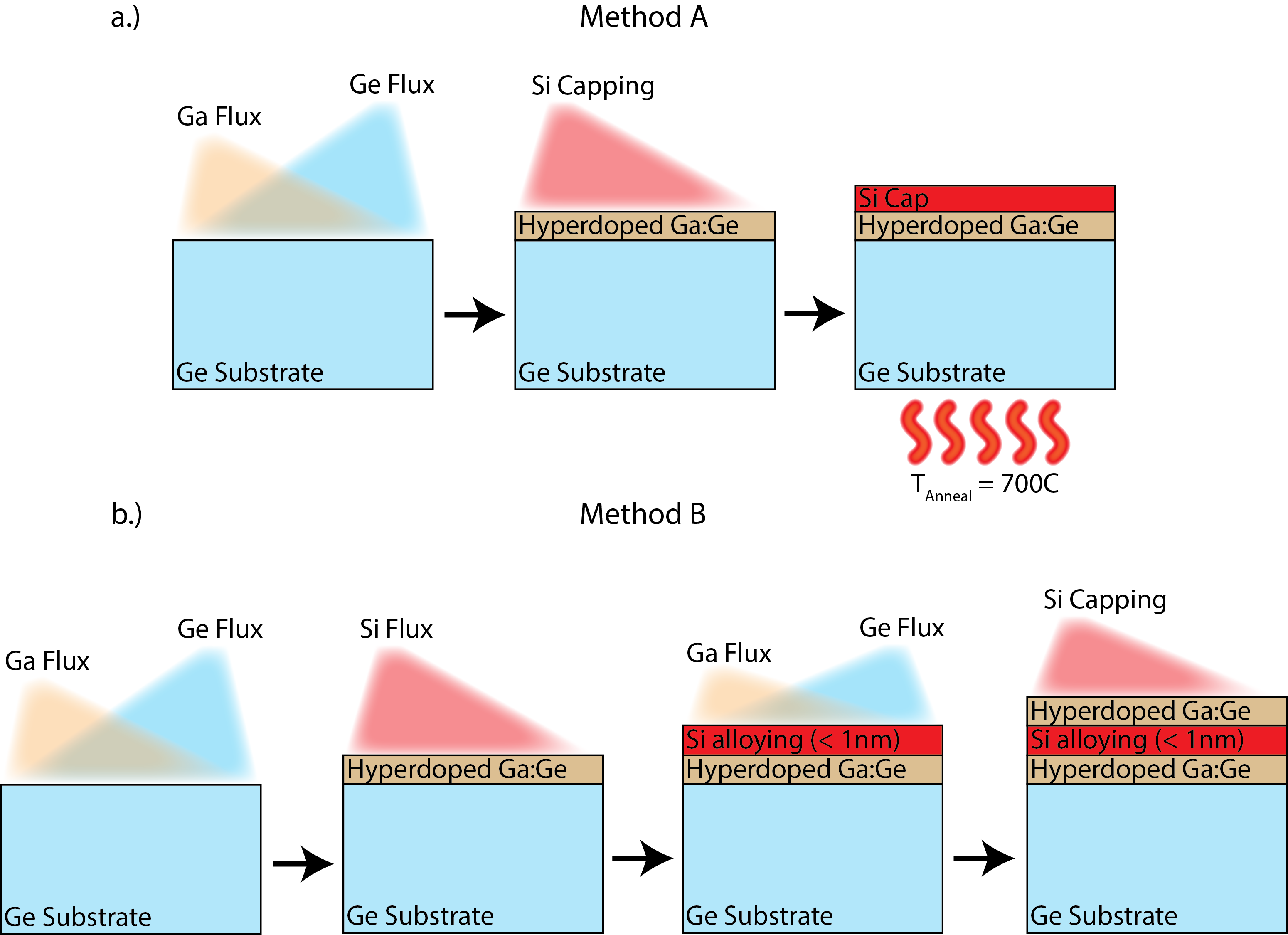}
    \caption{\textbf{a.)} Schematic of growth method A. We first co-deposit Ga and Ge species followed by a Si cap that we oxidize in the load lock. We then re-introduce the wafer into UHV and anneal at 700C. \textbf{b.)} Schematic of growth method B. Here, we co-deposit Ga and Ge species in two steps, separated by a thin layer of Si which we wish to alloy into the film. Lastly we cap with Si before removing from vacuum. No annealing is done with method B.}
    \label{fig1_mech-cartoon}
\end{figure}

Here we present the first report of molecular beam epitaxial (MBE) growth of superconducting hyperdoped Ga:Ge thin films. We show that under the appropriate growth conditions Ga-segregation is suppressed and a unique superconducting phase is obtained. We find distinctive signatures between superconductivity that is originated from Ga-metal as well as dispersive hyperdoping, schematically shown in Figure \ref{fig1_mech-cartoon}, through a combined electrical and structural analysis. 

%There are typically two extreme cases in Ga deposition techniques. Figure \ref{fig1_mech-cartoon}a shows schematically the presence of segregated Ga metal where we expect to observe ''puddles`` of Ga that are well-defined with respect to the surrounding Ge matrix. The interface contains numerous polymorphs of Ga, none of which crystallize in a well lattice-matched way to Ge. The Ga particle is expected to be either fully amorphous or polycrystalline. Furthermore, since the film is quite thin ($\sim$5-10~nm), we expect this segregation to occur entirely at the film-vacuum interface. Second, as pictured in Figure \ref{fig1_mech-cartoon}b, for films in which the hyperdoping maintains a dispersive nature with regards to Ga concentration we expect to see a coherent crystalline interface at the film-substrate interface. The film will also be fully connected, opposed to the percolated nature of the system pictured in Fig. \ref{fig1_mech-cartoon}a. 

The materials presented here were all grown in a custom Varian Gen II MBE chamber on 2" un-doped Ge (001) wafers. Prior to growth, wafers are etched \textit{ex-situ} in DI-water at 90$^{\circ}$C and then immediately loaded into vacuum. The wafers were then outgassed \textit{in-situ} at 400$^{\circ}$C for 30 minutes and then finally flash annealed at 650$^{\circ}$C for 5 minutes before cooling to the growth temperature. Pure Germanium and silicon are deposited via Thermionics HM2 e-Gun operating with a 10~kV acceleration voltage. Gallium doping is done with a standard knudsen cell source (MBE Komponenten) at a source temperature of 950$^{\circ}$C, or an approximate flux at the source opening of $\sim$1.5$\times$10$^{16}$/cm$^{2}$s as estimated by the kinetic theory of gases. The Ge source is operated at a 10kV acceleration voltage at 0.05A emission current while Si is evaporated with 0.01A emission current. Substrates are mounted in indium-free bayonet style holders and the reported substrate temperatures are measured with a thermocouple in close proximity to the backside of the wafer. The substrate is rotated at a constant 10 rpm throughout film growth to promote uniformity. 

In this study, we discuss two different sample growths and processing methods to help illuminate the Ga dopant atom behavior. The first method A seen in Figure \ref{fig1_mech-cartoon}a uses room temperature co-deposition of Ga and Ge onto the substrate followed by a capping layer of silicon under UHV. We oxidize the capping layer in the chamber load lock. The sample is then re-introduced to the growth chamber and flash annealed at 700$^{\circ}$ and cooled back down, emulating the processing conditions of ion implantation samples \cite{sardashti2021GaGeprm, herrmannsdorfer2009superGaGe, fiedler2012GaGeSiO2capping}. The second method B, schematically pictured in Figure \ref{fig1_mech-cartoon}b uses room temperature co-deposition of Ga and Ge, however halfway through the growth of the superconducting layer, growth is paused and a thin ~1nm thick layer of silicon is included before growth of Ga and Ge is continued. Deposition of Si at an intermediate stage of growth introduces a small amount of Si that alloys into the hyperdoped region, which is expected to increase the average phonon frequency, $\omega_{ln}$. As follows from the McMillan equation \cite{blase2009superIV, carbotte1990bosonxchange} seen in Equation \ref{mcmillanEQ}, we expect this to then give rise to an increased $T_{c}$, or superconducting transition temperature. We then cap the sample with an additional layer of silicon after completing the deposition of hyperdoped Ge. No post-annealing is performed for samples grown using method B.

\begin{equation}
    T_{c} = \frac{h\omega_{ln}}{1.2k_{B}}Exp\left[\frac{-1.04(1+\lambda_{ep})}{\lambda_{ep}-\mu^{*}(1+0.62\lambda_{ep})}\right]
    \label{mcmillanEQ}
\end{equation}

 Where $\lambda_{ep}$ is the electron-phonon coupling parameter and $\mu^{*}$ is the screened retarded Coulomb repulsion parameter. Growth method B attempts to utilize the McMillan formula \cite{blase2009superIV} in Eq.~\ref{mcmillanEQ} which suggests alloying with silicon to increase the average phonon frequency of the system ($\omega_{ln}$), thus enhancing the observed $T_{c}$ for a given density of states. Additionally, an increase in $\omega_{ln}$ is further expected to also increase the electron-phonon coupling potential, $V_{ep}$. By the relationship $V_{ep}=\lambda_{ep}/N(E_{F})$, where $N(E_{F})$ refers to the density of states at the Fermi level, we see the electron-phonon coupling parameter ($\lambda_{ep}$) increases with the density of states. Both of these factors could contribute to an enhanced $T_{c}$ \cite{carbotte1990bosonxchange}.

\begin{figure*}[ht!]
    \centering
    \includegraphics[width=\linewidth]{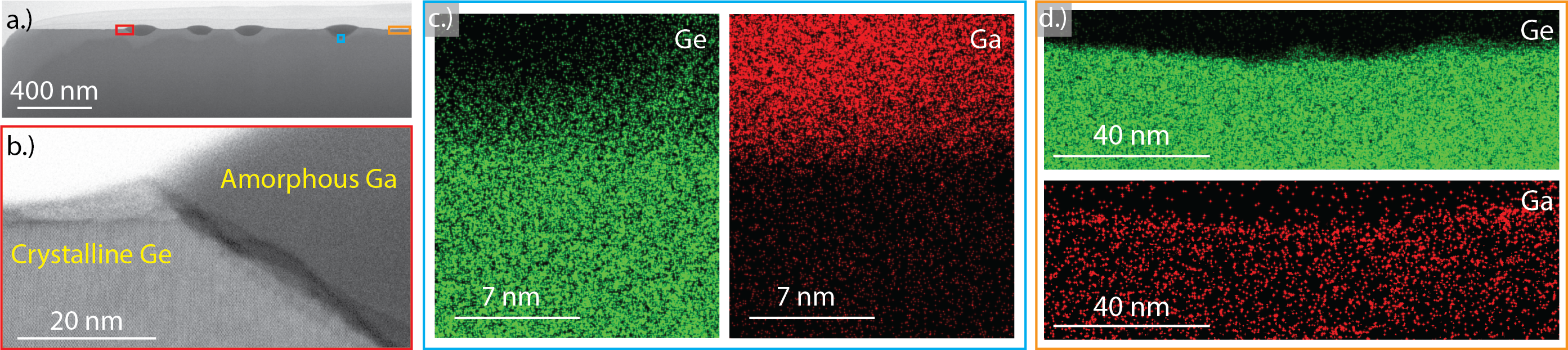}
    \caption{Bright field (BF) cross-sectional STEM images of a Ga-segregated Ge film. \textbf{(a)} Long-range scan showing significant segregation and agglomeration of Ga-droplets. \textbf{(b)} Zoom-in on region indicated in (a) showing the triple interface of epoxy(vacuum), Ga-droplet, and underlying Ge. \textbf{(c)} EDS map of region denoted by blue box in (a) of the interface between a droplet and the substrate. \textbf{(d)} EDS map of the region highlighted in the orange box in (a) of the region between Ga-rich droplets.}
    \label{fig_ga-seg}
\end{figure*}

The crystallinity, compositional distribution, and film morphology of the films grown in this study were examined with scanning transmission electron microscopy (STEM) in a JEOL ARM200F, equipped with a spherical aberration corrector for probe mode, and operated at 200 keV. The samples were prepared with cross-sectional tripod polishing to 20 $\mu$m thickness, followed by shallow angle Ar+ ion milling with low beam energies([1]3 keV), and LN2 stage cooling in a PIPS II ion mill. Cross-sectional STEM images of a sample grown using Method A are seen in Figure \ref{fig_ga-seg}a-b. Rapid annealing films under vacuum induces an extreme segregation of gallium metal out of the Ge matrix to form amorphous droplets on the surface of the substrate/film. These droplets selectively form in the near surface region and have a faint shadowing feature at the Ga-Ge interface which we attribute to the heavy Ga-content in that region. The droplets on average are hundreds of nm in diameter, with a similar spacing separating the droplets from one another. The droplets are not isotropic and extend further into the substrate than they protrude from the surface.

Zooming into the interface between the droplet and the underlying Ge, we confirm the amorphous nature of the Ga metal, as seen in Figure \ref{fig_ga-seg}b. This distinctly contrasts the crystalline nature of the underlying Ge. In this scenario, the Ga-rich regions exhibit significant out-diffusion of Ga metal which we speculate is likely due to the high temperature anneal used to activate the Ga-dopants. We further confirm this compositional shift through EDS maps, presented in Figure \ref{fig_ga-seg}c-d, of both a droplet region (blue) and the region between droplets (orange). Figure \ref{fig_ga-seg}c is zoomed in on the interface at the bottom of a droplet and the underlying Ge, depicted in Fig. \ref{fig_ga-seg}a as the small blue box. We see a $\sim$5~nm thick Ga-rich Ge region before the composition becomes nearly pure Ga. Between droplets, as seen in the area highlighted in the orange box in Fig. \ref{fig_ga-seg}a,  we instead see a very thin Ga-rich region on the order of few nanometers thick that serve as metallic interconnects between the Ga droplets. 

\begin{figure}[h!]
    \centering
    \includegraphics[width=\linewidth]{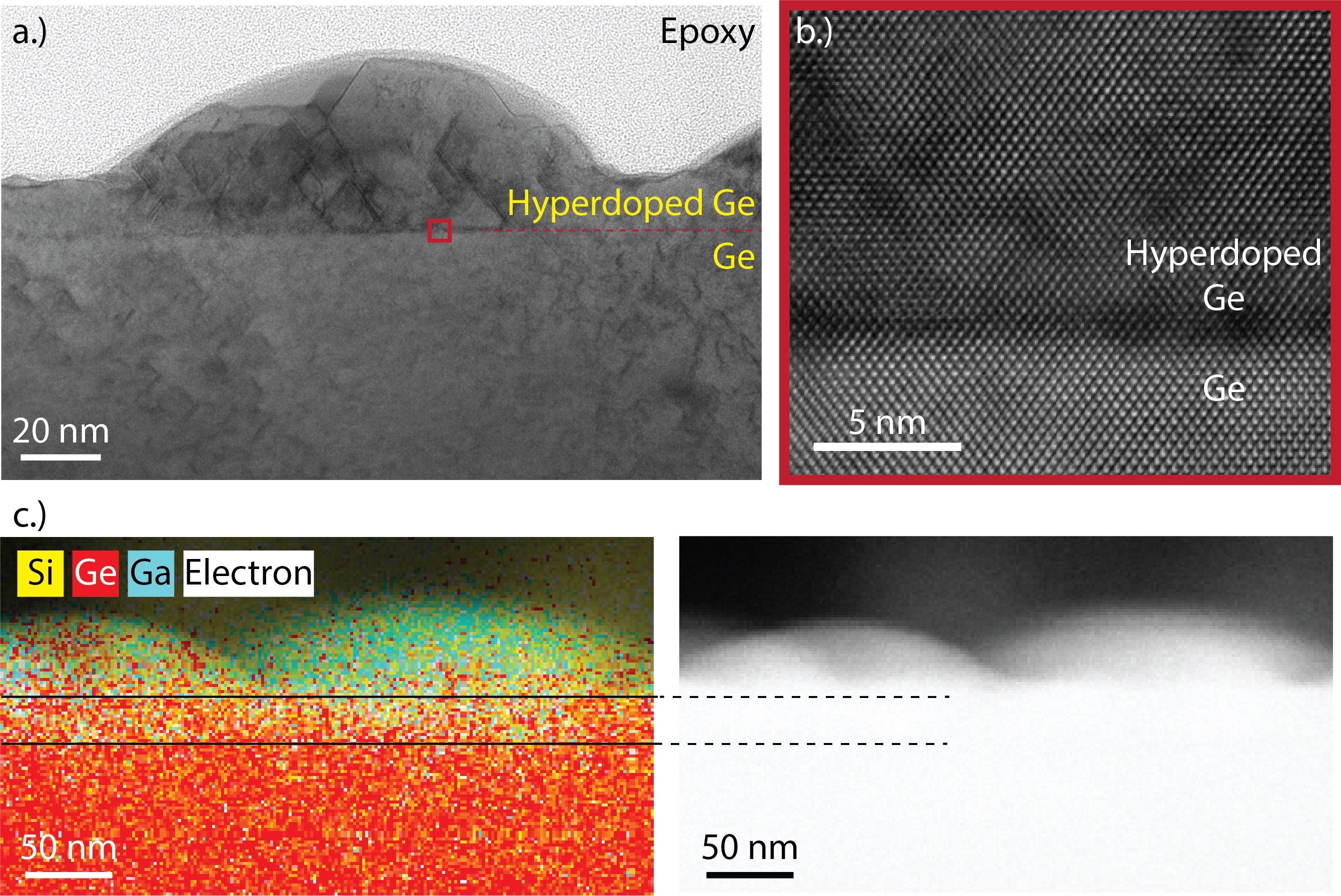}
    \caption{\textbf{(a)} High resolution TEM image to show the general film structure of a crystalline, hyperdoped Ge film. \textbf{(b)} Zoom-in on region indicated in (a) to present the epitaxial interface between the hyperdoped region and the underlying Ge buffer. \textbf{(c)} EDS elemental mapping of a zoomed out region showing elemental composition of the film and hillocks. Yellow pixels are Si, red is Ge, and cyan is Ga.}
    \label{fig_HD-ge}
\end{figure}

To suppress this Ga metal segregation, in our method B we deposit the heavily doped layers at room temperature, however this time we do not anneal at elevated temperatures. Removing the anneal step allows for maintaining a nominally consistent film composition and eliminate the observed Ga metal precipitates. Such behavior is pictured in Figure \ref{fig_HD-ge} in which cross-sectional STEM images for a sample grown using Method B present a fully connected film with an abrupt film/substrate interface.

In the second method in which we utilize no post-anneal for carrier activation, Fig. \ref{fig_HD-ge} shows a nominally sharp interface between the heavily-doped region and the undoped germanium. Fig. \ref{fig_HD-ge}a shows a fully complete film coverage with no obvious signatures of significant Ga segregation, albeit the film is quite rough. We attribute film roughness to the difference in surface energy between Si and Ge giving rise to a 3D island growth mechanism of Si on Ge during the Si cap growth. This behavior of the Si-Ge system has been previously observed and reported on \cite{tsu1994siongembe, kawabata1989siheteroepi, copel1989episurfactant}. Furthermore, as we zoom in on the interface, we observe no obvious discontinuities in the atomic columns suggesting the Ge lattice is maintained as Ga has been incorporated. 

Fig. \ref{fig_HD-ge}c shows the EDS elemental mapping of a zoomed out region showing elemental composition of the film. The interface between Ge and doped Ge is marked. The effective thickness of the doped Ge film is 5~nm before hillocks. The map also shows that hilllocks are predominantly Ga-doped Si while in some places Ge is present. The continuous atomic registry is highly promising for continued development of coherent superconducting germanium thin films.

Electrical measurements are conducted in an Oxford Triton pulse-tube dilution refrigerator with a base temperature of 15~mK and magnetic field capabilities up to 14T. Measurements are collected using a standard Van der Pauw wiring configuration on square pieces from near the center of each wafer. On-chip contacts are annealed In-Sn eutectic at each of the four corners.

\begin{figure}[h!]
    \centering
    \includegraphics[width=\linewidth]{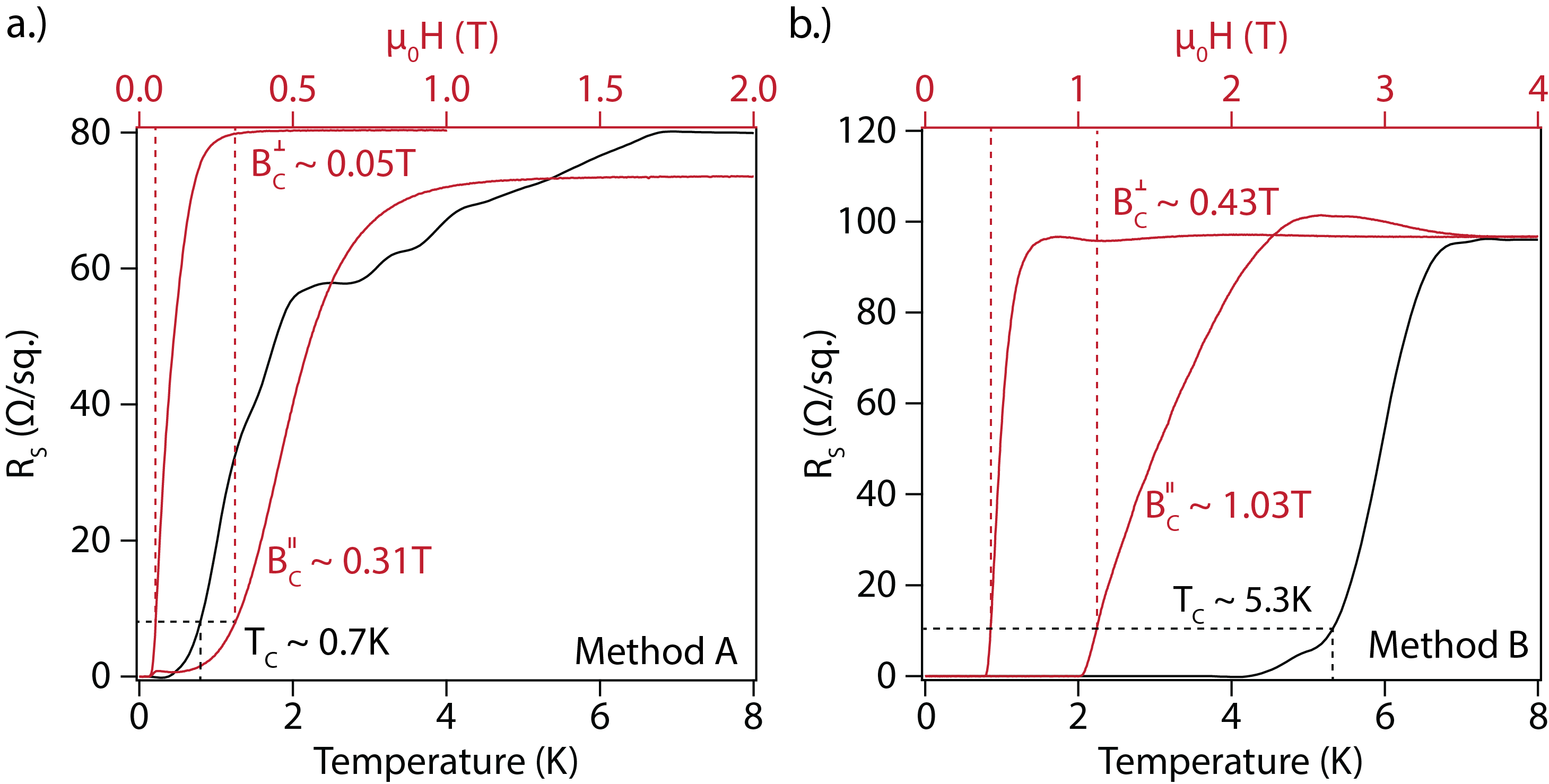}
    \caption{\textbf{a.)} Sheet resistance vs. temperature (black) and magnetic field (red, at 15mK) of a sample grown using Method A. Critical transition temperature is near 0.7K, with a critical out-of-plane magnetic field of roughly 50mT and critical in-plane magnetic field of $\sim$310mT. \textbf{b.)} Sheet resistance vs temperature (black) and magnetic field (red, at 15mK) of a sample grown using Method B. Sample B exhibits a transition temperature of near 5.3K, with an enhanced critical out-of-plane field of 430mT and critial in-plane field of $\sim$1.03T.}
    \label{fig_transport}
\end{figure}

 Transport measurements for both methods are presented in Figure \ref{fig_transport}. In our Ga-segregated films, grown via method A, the observed critical temperature and of 0.89~K at zero field, and critical field values of $B_{c}^{\perp} \sim 0.05~T$, and $B_{c}^{\parallel} \sim 310mT$ at T = 15~mK, are highly suggestive of a Ga-metal origin. The transitions are marked by the 10\% value of the normal resistance. Reported literature values \cite{eisenstein1954gasupercond} for Ga of 1.1K and 0.05T, respectively, agree well with our measurements. The observed reduction in critical temperature is most likely due to the Ga-droplets behaving as a weakly connected superconductor such that discrete ``puddles'' of superconducting Ga metal host the parent superconducting phase, but due to the low density we do not observe exactly the same T$_{c}$ as bulk Ga. The enhanced in-plane field of 0.31~T we attribute to the thin film nature of the superconducting film \cite{Tedrow73}.

 Looking closer at the sheet resistance as a function of temperature, many kinks are observed between the range of $\sim$7K and the total superconducting transition at 0.7K. We attribute the high temperature kink to the formation of Ga metal crystalline polymorphs as a result of the anneal, all of which have been shown previously to exhibit a superconducting transition temperature of 6K or greater \cite{charnaya1998confinedga, charnaya2009gananoconfine, moura2017betaGa, teske1999gapolymorph}. The other features that are presented at more moderate temperatures $\sim$2-4K are more difficult to confidently assign to Ga-related phases, but could be due to either percolated Ga metal networks or sparse regions of hyperdoped Ge matrix.

Fig. \ref{fig_transport}b shows a different trend for samples grown using Method B. Here, we observe significantly enhanced superconducting properties compared to that of the Ga-segregated films with $T_{C} = 5.27K$, $B_{c}^{\perp} = 0.43T$, and $B_{c}^{\parallel} = 1.03T$. Here, in the sheet resistance versus temperature we observe a much sharper transition that starts $\sim$6.9K and reaches the zero resistance state at $\sim$5K. Previous reports of superconductivity in Ga metal polymorphs have observed superconductivity at 6.9K in $\epsilon$-Ge and $\beta$-Ga at 6K \cite{teske1999gapolymorph, moura2017betaGa}. Unfortunately, magnetic field behavior for many of these crystal polymorphs of Ga is not reported in literature, however reported values for nanoconfined particles of $\beta$-Ga are on the order of $\sim$430~Oe \cite{moura2017betaGa}. While longer-extent TEM images do show a single small $\beta$-Ga nanoparticle present in the FIB slice, the observed out-of-plane critical magnetic field exhibited by our hyperdoped Ge film is significantly larger than any reported values for crystalline phases of Ga. Thus, we are confident in ruling out competing Ga phases as the origin of the observed superconductivity due to their sparsity in our film. If the superconductivity observed here were a result of sparse interconnected grains of some alternative phase of Ga, we would expect a much larger transitional region, similar to what we observe for the sample grown via method A and as has been reported previously in studies on Nb islands \cite{eley2011nbisland}.

%Comparing the field-dependant behavior seen in the hyperdoped Ge sample (method B), we observe a broad transitional region with in-plane field suggestive of vortex motion associated with type II superconductivity \cite{tinkham2004introduction}. With an out-of-plane field orientation however, we observe a much sharper transition out of the superconducting state.

To help understand the superconducting phase in the sample grown using method B, we conduct Hall measurements, above $B_{c}$, in the Van der Pauw geometry and compare against the theorized quantities for an electronic origin. We measure a 2D hole concentration of roughly 2.86$\times$10$^{12}$/cm$^{2}$ for the specific sample presented in this work, or a density of $\sim$1$\times$10$^{19}$/$cm^{3}$ carriers. Comparing to previous reports of covalent superconductivity in this and related systems \cite{blase2009superIV, bustarret2006superSilicon, ekimov2004superDiamond, herrmannsdorfer2009superGaGe}, we see that not only is the transition temperature observed in this study significantly higher than is predicted for the Ge system ($\sim$100s mK) \cite{cohen1964supersemi, cohen1964supersemi2}, but we also measure drastically lower carrier densities. 

In conclusion, we have presented here our report of superconductivity in Ga-doped Germanium achieved via molecular beam epitaxy. The superconducting phase is sensitive to processing conditions and we observe that high temperature flash anneals strongly promote Ga-segregation in the matrix. Furthermore, we find that the MBE deposition process enables carrier activation even without a post-anneal in which superconductivity is attained. The hyperdoped film we presented here exhibits a critical temperature of $T_{C} = 5.27K$ and critical field values of $B_{C}^{\perp} = 0.43T, B_{C}^{\parallel}=1.03T$ at 15~mK. We lastly found that the interface between un-doped Ge and the hyperdoped Ge layer remains highly coherent after growth. Therefore the results shown here present MBE-growth of superconducting group IV material as a highly promising approach for the integration of superconducting Ge films, though much more work is required to fully realize full group IV superconducting circuitry.

\section*{Author Declarations}
\subsection*{Conflict of Interest}
The authors have no conflicts to disclose.

\section*{Author's Contributions}
\noindent \textbf{Patrick J. Strohbeen: } Conceptualization (equal); Data Curation (lead); Formal Analysis (lead); Investigation (lead); Methodology (equal); Project Administration (supporting); Software (supporting); Supervision (equal); Validation (equal); Visualization (lead); Writing - Original Draft (lead); Writing - Review \& Editing (equal); 
\textbf{Aurelia M. Brook: } Formal Analysis (supporting); Investigation (supporting); Software (lead); Visualization (supporting); Writing - Review \& Editing (equal);
\textbf{Wendy L. Sarney: } Investigation (supporting); Writing - Review \& Editing (supporting);
\textbf{Javad Shabani: } Conceptualization (equal); Funding Acquisition (lead); Methodology (equal); Project Administration (lead); Supervision (equal); Validation (equal); Writing - Review \& Editing (equal); 

\begin{acknowledgments}
This project was supported by AFOSR award FA9550-21-1-0338.

\end{acknowledgments}

\section*{Data Availability Statement}

The data that support the findings of this study are available from the corresponding author upon reasonable request.

\nocite{*}
\bibliography{bibliography}% Produces the bibliography via BibTeX.

\begin{thebibliography}{29}%
\makeatletter
\providecommand \@ifxundefined [1]{%
 \@ifx{#1\undefined}
}%
\providecommand \@ifnum [1]{%
 \ifnum #1\expandafter \@firstoftwo
 \else \expandafter \@secondoftwo
 \fi
}%
\providecommand \@ifx [1]{%
 \ifx #1\expandafter \@firstoftwo
 \else \expandafter \@secondoftwo
 \fi
}%
\providecommand \natexlab [1]{#1}%
\providecommand \enquote  [1]{``#1''}%
\providecommand \bibnamefont  [1]{#1}%
\providecommand \bibfnamefont [1]{#1}%
\providecommand \citenamefont [1]{#1}%
\providecommand \href@noop [0]{\@secondoftwo}%
\providecommand \href [0]{\begingroup \@sanitize@url \@href}%
\providecommand \@href[1]{\@@startlink{#1}\@@href}%
\providecommand \@@href[1]{\endgroup#1\@@endlink}%
\providecommand \@sanitize@url [0]{\catcode `\\12\catcode `\$12\catcode
  `\&12\catcode `\#12\catcode `\^12\catcode `\_12\catcode `\%12\relax}%
\providecommand \@@startlink[1]{}%
\providecommand \@@endlink[0]{}%
\providecommand \url  [0]{\begingroup\@sanitize@url \@url }%
\providecommand \@url [1]{\endgroup\@href {#1}{\urlprefix }}%
\providecommand \urlprefix  [0]{URL }%
\providecommand \Eprint [0]{\href }%
\providecommand \doibase [0]{http://dx.doi.org/}%
\providecommand \selectlanguage [0]{\@gobble}%
\providecommand \bibinfo  [0]{\@secondoftwo}%
\providecommand \bibfield  [0]{\@secondoftwo}%
\providecommand \translation [1]{[#1]}%
\providecommand \BibitemOpen [0]{}%
\providecommand \bibitemStop [0]{}%
\providecommand \bibitemNoStop [0]{.\EOS\space}%
\providecommand \EOS [0]{\spacefactor3000\relax}%
\providecommand \BibitemShut  [1]{\csname bibitem#1\endcsname}%
\let\auto@bib@innerbib\@empty
%</preamble>
\bibitem [{\citenamefont {Patra}\ \emph {et~al.}(2018)\citenamefont {Patra},
  \citenamefont {Incandela}, \citenamefont {van Dijk}, \citenamefont {Homulle},
  \citenamefont {Song}, \citenamefont {Shahmohammadi}, \citenamefont
  {Staszewski}, \citenamefont {Vladimirescu}, \citenamefont {Babaie},
  \citenamefont {Sebastiano},\ and\ \citenamefont
  {Charbon}}]{patra2017cryocmos}%
  \BibitemOpen
  \bibfield  {author} {\bibinfo {author} {\bibfnamefont {B.}~\bibnamefont
  {Patra}}, \bibinfo {author} {\bibfnamefont {R.~M.}\ \bibnamefont
  {Incandela}}, \bibinfo {author} {\bibfnamefont {J.~P.~G.}\ \bibnamefont {van
  Dijk}}, \bibinfo {author} {\bibfnamefont {H.~A.~R.}\ \bibnamefont {Homulle}},
  \bibinfo {author} {\bibfnamefont {L.}~\bibnamefont {Song}}, \bibinfo {author}
  {\bibfnamefont {M.}~\bibnamefont {Shahmohammadi}}, \bibinfo {author}
  {\bibfnamefont {R.~B.}\ \bibnamefont {Staszewski}}, \bibinfo {author}
  {\bibfnamefont {A.}~\bibnamefont {Vladimirescu}}, \bibinfo {author}
  {\bibfnamefont {M.}~\bibnamefont {Babaie}}, \bibinfo {author} {\bibfnamefont
  {F.}~\bibnamefont {Sebastiano}}, \ and\ \bibinfo {author} {\bibfnamefont
  {E.}~\bibnamefont {Charbon}},\ }\href@noop {} {\bibfield  {journal} {\bibinfo
   {journal} {IEEE J. Solid-State Circuits}\ }\textbf {\bibinfo {volume}
  {53}},\ \bibinfo {pages} {309} (\bibinfo {year} {2018})}\BibitemShut
  {NoStop}%
\bibitem [{\citenamefont {Shim}\ and\ \citenamefont {Tahan}(2016)}]{ShimTahan}%
  \BibitemOpen
  \bibfield  {author} {\bibinfo {author} {\bibfnamefont {Y.-P.}\ \bibnamefont
  {Shim}}\ and\ \bibinfo {author} {\bibfnamefont {C.}~\bibnamefont {Tahan}},\
  }\href@noop {} {\enquote {\bibinfo {title} {Semiconductor-inspired design
  principles for superconducting quantum computing},}\ } (\bibinfo {year}
  {2016})\BibitemShut {NoStop}%
\bibitem [{\citenamefont {Zhao}\ \emph {et~al.}(2020)\citenamefont {Zhao},
  \citenamefont {Park}, \citenamefont {Zhao}, \citenamefont {Bal},
  \citenamefont {McRae}, \citenamefont {Long},\ and\ \citenamefont
  {Pappas}}]{zhao2020merge}%
  \BibitemOpen
  \bibfield  {author} {\bibinfo {author} {\bibfnamefont {R.}~\bibnamefont
  {Zhao}}, \bibinfo {author} {\bibfnamefont {S.}~\bibnamefont {Park}}, \bibinfo
  {author} {\bibfnamefont {T.}~\bibnamefont {Zhao}}, \bibinfo {author}
  {\bibfnamefont {M.}~\bibnamefont {Bal}}, \bibinfo {author} {\bibfnamefont
  {C.}~\bibnamefont {McRae}}, \bibinfo {author} {\bibfnamefont
  {J.}~\bibnamefont {Long}}, \ and\ \bibinfo {author} {\bibfnamefont {D.~P.}\
  \bibnamefont {Pappas}},\ }\href@noop {} {\bibfield  {journal} {\bibinfo
  {journal} {Phys. Rev. Applied}\ }\textbf {\bibinfo {volume} {14}},\ \bibinfo
  {pages} {064006} (\bibinfo {year} {2020})}\BibitemShut {NoStop}%
\bibitem [{\citenamefont {Shim}\ and\ \citenamefont
  {Tahan}(2014)}]{tahan2014bottomup}%
  \BibitemOpen
  \bibfield  {author} {\bibinfo {author} {\bibfnamefont {Y.-P.}\ \bibnamefont
  {Shim}}\ and\ \bibinfo {author} {\bibfnamefont {C.}~\bibnamefont {Tahan}},\
  }\href@noop {} {\bibfield  {journal} {\bibinfo  {journal} {Nat. Commun.}\
  }\textbf {\bibinfo {volume} {5}},\ \bibinfo {pages} {4225} (\bibinfo {year}
  {2014})}\BibitemShut {NoStop}%
\bibitem [{\citenamefont {McRae}\ \emph {et~al.}(2021)\citenamefont {McRae},
  \citenamefont {McFadden}, \citenamefont {Zhao}, \citenamefont {Wang},
  \citenamefont {Long}, \citenamefont {Zhao}, \citenamefont {Park},
  \citenamefont {Bal}, \citenamefont {Palmstr{\o}m},\ and\ \citenamefont
  {Pappas}}]{mcrae2021IIIVMerge}%
  \BibitemOpen
  \bibfield  {author} {\bibinfo {author} {\bibfnamefont {C.~R.~H.}\
  \bibnamefont {McRae}}, \bibinfo {author} {\bibfnamefont {A.}~\bibnamefont
  {McFadden}}, \bibinfo {author} {\bibfnamefont {R.}~\bibnamefont {Zhao}},
  \bibinfo {author} {\bibfnamefont {H.}~\bibnamefont {Wang}}, \bibinfo {author}
  {\bibfnamefont {J.~L.}\ \bibnamefont {Long}}, \bibinfo {author}
  {\bibfnamefont {T.}~\bibnamefont {Zhao}}, \bibinfo {author} {\bibfnamefont
  {S.}~\bibnamefont {Park}}, \bibinfo {author} {\bibfnamefont {M.}~\bibnamefont
  {Bal}}, \bibinfo {author} {\bibfnamefont {C.~J.}\ \bibnamefont
  {Palmstr{\o}m}}, \ and\ \bibinfo {author} {\bibfnamefont {D.~P.}\
  \bibnamefont {Pappas}},\ }\href@noop {} {\bibfield  {journal} {\bibinfo
  {journal} {J. Appl. Phys.}\ }\textbf {\bibinfo {volume} {129}},\ \bibinfo
  {pages} {025109} (\bibinfo {year} {2021})}\BibitemShut {NoStop}%
\bibitem [{\citenamefont {Nakamura}\ \emph {et~al.}(1999)\citenamefont
  {Nakamura}, \citenamefont {Pashkin},\ and\ \citenamefont
  {Tsai}}]{nakamura1999cpbox}%
  \BibitemOpen
  \bibfield  {author} {\bibinfo {author} {\bibfnamefont {Y.}~\bibnamefont
  {Nakamura}}, \bibinfo {author} {\bibfnamefont {Y.~A.}\ \bibnamefont
  {Pashkin}}, \ and\ \bibinfo {author} {\bibfnamefont {J.~S.}\ \bibnamefont
  {Tsai}},\ }\href@noop {} {\bibfield  {journal} {\bibinfo  {journal} {Nature}\
  }\textbf {\bibinfo {volume} {398}},\ \bibinfo {pages} {786} (\bibinfo {year}
  {1999})}\BibitemShut {NoStop}%
\bibitem [{\citenamefont {Vion}\ \emph {et~al.}(2002)\citenamefont {Vion},
  \citenamefont {Aassime}, \citenamefont {Cottet}, \citenamefont {Joyez},
  \citenamefont {Pothier}, \citenamefont {Urbina}, \citenamefont {Esteve},\
  and\ \citenamefont {Devoret}}]{vion2002quantronium}%
  \BibitemOpen
  \bibfield  {author} {\bibinfo {author} {\bibfnamefont {D.}~\bibnamefont
  {Vion}}, \bibinfo {author} {\bibfnamefont {A.}~\bibnamefont {Aassime}},
  \bibinfo {author} {\bibfnamefont {A.}~\bibnamefont {Cottet}}, \bibinfo
  {author} {\bibfnamefont {P.}~\bibnamefont {Joyez}}, \bibinfo {author}
  {\bibfnamefont {H.}~\bibnamefont {Pothier}}, \bibinfo {author} {\bibfnamefont
  {C.}~\bibnamefont {Urbina}}, \bibinfo {author} {\bibfnamefont
  {D.}~\bibnamefont {Esteve}}, \ and\ \bibinfo {author} {\bibfnamefont {M.~H.}\
  \bibnamefont {Devoret}},\ }\href@noop {} {\bibfield  {journal} {\bibinfo
  {journal} {Science}\ }\textbf {\bibinfo {volume} {296}},\ \bibinfo {pages}
  {886} (\bibinfo {year} {2002})}\BibitemShut {NoStop}%
\bibitem [{\citenamefont {Ekimov}\ \emph {et~al.}(2004)\citenamefont {Ekimov},
  \citenamefont {Sidorov}, \citenamefont {Bauer}, \citenamefont {Mel'nik},
  \citenamefont {Curro}, \citenamefont {Thompson},\ and\ \citenamefont
  {Stishov}}]{ekimov2004superDiamond}%
  \BibitemOpen
  \bibfield  {author} {\bibinfo {author} {\bibfnamefont {E.~A.}\ \bibnamefont
  {Ekimov}}, \bibinfo {author} {\bibfnamefont {V.~A.}\ \bibnamefont {Sidorov}},
  \bibinfo {author} {\bibfnamefont {E.~D.}\ \bibnamefont {Bauer}}, \bibinfo
  {author} {\bibfnamefont {N.~N.}\ \bibnamefont {Mel'nik}}, \bibinfo {author}
  {\bibfnamefont {N.~J.}\ \bibnamefont {Curro}}, \bibinfo {author}
  {\bibfnamefont {J.~D.}\ \bibnamefont {Thompson}}, \ and\ \bibinfo {author}
  {\bibfnamefont {S.~M.}\ \bibnamefont {Stishov}},\ }\href@noop {} {\bibfield
  {journal} {\bibinfo  {journal} {Nature}\ }\textbf {\bibinfo {volume} {428}},\
  \bibinfo {pages} {542} (\bibinfo {year} {2004})}\BibitemShut {NoStop}%
\bibitem [{\citenamefont {Bustarret}\ \emph {et~al.}(2006)\citenamefont
  {Bustarret}, \citenamefont {Marcenat}, \citenamefont {Achatz}, \citenamefont
  {Ka\v{c}mar\v{c}ik}, \citenamefont {L\'{e}vy}, \citenamefont {Huxley},
  \citenamefont {Ort\'{e}ga}, \citenamefont {Bourgeois}, \citenamefont {Blase},
  \citenamefont {D\'{e}barre},\ and\ \citenamefont
  {Boulmer}}]{bustarret2006superSilicon}%
  \BibitemOpen
  \bibfield  {author} {\bibinfo {author} {\bibfnamefont {E.}~\bibnamefont
  {Bustarret}}, \bibinfo {author} {\bibfnamefont {C.}~\bibnamefont {Marcenat}},
  \bibinfo {author} {\bibfnamefont {P.}~\bibnamefont {Achatz}}, \bibinfo
  {author} {\bibfnamefont {J.}~\bibnamefont {Ka\v{c}mar\v{c}ik}}, \bibinfo
  {author} {\bibfnamefont {F.}~\bibnamefont {L\'{e}vy}}, \bibinfo {author}
  {\bibfnamefont {A.}~\bibnamefont {Huxley}}, \bibinfo {author} {\bibfnamefont
  {L.}~\bibnamefont {Ort\'{e}ga}}, \bibinfo {author} {\bibfnamefont
  {E.}~\bibnamefont {Bourgeois}}, \bibinfo {author} {\bibfnamefont
  {X.}~\bibnamefont {Blase}}, \bibinfo {author} {\bibfnamefont
  {D.}~\bibnamefont {D\'{e}barre}}, \ and\ \bibinfo {author} {\bibfnamefont
  {J.}~\bibnamefont {Boulmer}},\ }\href@noop {} {\bibfield  {journal} {\bibinfo
   {journal} {Nature}\ }\textbf {\bibinfo {volume} {444}},\ \bibinfo {pages}
  {465} (\bibinfo {year} {2006})}\BibitemShut {NoStop}%
\bibitem [{\citenamefont {Sardashti}\ \emph
  {et~al.}(2021{\natexlab{a}})\citenamefont {Sardashti}, \citenamefont
  {Nguyen}, \citenamefont {Hatefipour}, \citenamefont {Sarney}, \citenamefont
  {Yuan}, \citenamefont {Mayer}, \citenamefont {Kisslinger},\ and\
  \citenamefont {Shabani}}]{sardashti2021supersi}%
  \BibitemOpen
  \bibfield  {author} {\bibinfo {author} {\bibfnamefont {K.}~\bibnamefont
  {Sardashti}}, \bibinfo {author} {\bibfnamefont {T.}~\bibnamefont {Nguyen}},
  \bibinfo {author} {\bibfnamefont {M.}~\bibnamefont {Hatefipour}}, \bibinfo
  {author} {\bibfnamefont {W.~L.}\ \bibnamefont {Sarney}}, \bibinfo {author}
  {\bibfnamefont {J.}~\bibnamefont {Yuan}}, \bibinfo {author} {\bibfnamefont
  {W.}~\bibnamefont {Mayer}}, \bibinfo {author} {\bibfnamefont
  {K.}~\bibnamefont {Kisslinger}}, \ and\ \bibinfo {author} {\bibfnamefont
  {J.}~\bibnamefont {Shabani}},\ }\href@noop {} {\bibfield  {journal} {\bibinfo
   {journal} {Appl. Phys. Lett.}\ }\textbf {\bibinfo {volume} {118}},\ \bibinfo
  {pages} {073102} (\bibinfo {year} {2021}{\natexlab{a}})}\BibitemShut
  {NoStop}%
\bibitem [{\citenamefont {Sardashti}\ \emph
  {et~al.}(2021{\natexlab{b}})\citenamefont {Sardashti}, \citenamefont
  {Nguyen}, \citenamefont {Sarney}, \citenamefont {Leff}, \citenamefont
  {Hatefipour}, \citenamefont {Dartiailh}, \citenamefont {Yuan}, \citenamefont
  {Mayer},\ and\ \citenamefont {Shabani}}]{sardashti2021GaGeprm}%
  \BibitemOpen
  \bibfield  {author} {\bibinfo {author} {\bibfnamefont {K.}~\bibnamefont
  {Sardashti}}, \bibinfo {author} {\bibfnamefont {T.~D.}\ \bibnamefont
  {Nguyen}}, \bibinfo {author} {\bibfnamefont {W.~L.}\ \bibnamefont {Sarney}},
  \bibinfo {author} {\bibfnamefont {A.~C.}\ \bibnamefont {Leff}}, \bibinfo
  {author} {\bibfnamefont {M.}~\bibnamefont {Hatefipour}}, \bibinfo {author}
  {\bibfnamefont {M.~C.}\ \bibnamefont {Dartiailh}}, \bibinfo {author}
  {\bibfnamefont {J.}~\bibnamefont {Yuan}}, \bibinfo {author} {\bibfnamefont
  {W.}~\bibnamefont {Mayer}}, \ and\ \bibinfo {author} {\bibfnamefont
  {J.}~\bibnamefont {Shabani}},\ }\href@noop {} {\bibfield  {journal} {\bibinfo
   {journal} {Phys. Rev. Mater.}\ }\textbf {\bibinfo {volume} {5}},\ \bibinfo
  {pages} {064802} (\bibinfo {year} {2021}{\natexlab{b}})}\BibitemShut
  {NoStop}%
\bibitem [{\citenamefont {Fiedler}\ \emph {et~al.}(2012)\citenamefont
  {Fiedler}, \citenamefont {Heera}, \citenamefont {Skrotzki}, \citenamefont
  {Herrmannsd{\"o}rfer}, \citenamefont {Voelskow}, \citenamefont
  {M{\"u}cklich}, \citenamefont {Facsko}, \citenamefont {Reuther},
  \citenamefont {Perego}, \citenamefont {Heinig}, \citenamefont {Schmidt},
  \citenamefont {Skorupa}, \citenamefont {Gobsch},\ and\ \citenamefont
  {Helm}}]{fiedler2012GaGeSiO2capping}%
  \BibitemOpen
  \bibfield  {author} {\bibinfo {author} {\bibfnamefont {J.}~\bibnamefont
  {Fiedler}}, \bibinfo {author} {\bibfnamefont {V.}~\bibnamefont {Heera}},
  \bibinfo {author} {\bibfnamefont {R.}~\bibnamefont {Skrotzki}}, \bibinfo
  {author} {\bibfnamefont {T.}~\bibnamefont {Herrmannsd{\"o}rfer}}, \bibinfo
  {author} {\bibfnamefont {M.}~\bibnamefont {Voelskow}}, \bibinfo {author}
  {\bibfnamefont {A.}~\bibnamefont {M{\"u}cklich}}, \bibinfo {author}
  {\bibfnamefont {S.}~\bibnamefont {Facsko}}, \bibinfo {author} {\bibfnamefont
  {H.}~\bibnamefont {Reuther}}, \bibinfo {author} {\bibfnamefont
  {M.}~\bibnamefont {Perego}}, \bibinfo {author} {\bibfnamefont {K.-H.}\
  \bibnamefont {Heinig}}, \bibinfo {author} {\bibfnamefont {B.}~\bibnamefont
  {Schmidt}}, \bibinfo {author} {\bibfnamefont {W.}~\bibnamefont {Skorupa}},
  \bibinfo {author} {\bibfnamefont {G.}~\bibnamefont {Gobsch}}, \ and\ \bibinfo
  {author} {\bibfnamefont {M.}~\bibnamefont {Helm}},\ }\href@noop {} {\bibfield
   {journal} {\bibinfo  {journal} {Phys. Rev. B}\ }\textbf {\bibinfo {volume}
  {85}},\ \bibinfo {pages} {134530} (\bibinfo {year} {2012})}\BibitemShut
  {NoStop}%
\bibitem [{\citenamefont {Herrmannsd{\"o}rfer}\ \emph
  {et~al.}(2009)\citenamefont {Herrmannsd{\"o}rfer}, \citenamefont {Heera},
  \citenamefont {Ignatchik}, \citenamefont {Uhlarz}, \citenamefont
  {M{\"u}cklich}, \citenamefont {Posselt}, \citenamefont {Reuther},
  \citenamefont {Schmidt}, \citenamefont {Heinig}, \citenamefont {Skorupa},
  \citenamefont {Voelskow}, \citenamefont {W{\"u}ndisch}, \citenamefont
  {Skrotzki}, \citenamefont {Helm},\ and\ \citenamefont
  {Wosnitza}}]{herrmannsdorfer2009superGaGe}%
  \BibitemOpen
  \bibfield  {author} {\bibinfo {author} {\bibfnamefont {T.}~\bibnamefont
  {Herrmannsd{\"o}rfer}}, \bibinfo {author} {\bibfnamefont {V.}~\bibnamefont
  {Heera}}, \bibinfo {author} {\bibfnamefont {O.}~\bibnamefont {Ignatchik}},
  \bibinfo {author} {\bibfnamefont {M.}~\bibnamefont {Uhlarz}}, \bibinfo
  {author} {\bibfnamefont {A.}~\bibnamefont {M{\"u}cklich}}, \bibinfo {author}
  {\bibfnamefont {M.}~\bibnamefont {Posselt}}, \bibinfo {author} {\bibfnamefont
  {H.}~\bibnamefont {Reuther}}, \bibinfo {author} {\bibfnamefont
  {B.}~\bibnamefont {Schmidt}}, \bibinfo {author} {\bibfnamefont {K.-H.}\
  \bibnamefont {Heinig}}, \bibinfo {author} {\bibfnamefont {W.}~\bibnamefont
  {Skorupa}}, \bibinfo {author} {\bibfnamefont {M.}~\bibnamefont {Voelskow}},
  \bibinfo {author} {\bibfnamefont {C.}~\bibnamefont {W{\"u}ndisch}}, \bibinfo
  {author} {\bibfnamefont {R.}~\bibnamefont {Skrotzki}}, \bibinfo {author}
  {\bibfnamefont {M.}~\bibnamefont {Helm}}, \ and\ \bibinfo {author}
  {\bibfnamefont {J.}~\bibnamefont {Wosnitza}},\ }\href@noop {} {\bibfield
  {journal} {\bibinfo  {journal} {Phys. Rev. Lett.}\ }\textbf {\bibinfo
  {volume} {102}},\ \bibinfo {pages} {217003} (\bibinfo {year}
  {2009})}\BibitemShut {NoStop}%
\bibitem [{\citenamefont {Skrotzki}\ \emph {et~al.}(2011)\citenamefont
  {Skrotzki}, \citenamefont {Herrmannsd{\"o}rfer}, \citenamefont {Heera},
  \citenamefont {Fiedler}, \citenamefont {M{\"u}cklich}, \citenamefont {Helm},\
  and\ \citenamefont {Wosnitza}}]{skrotzki2011Gedoping}%
  \BibitemOpen
  \bibfield  {author} {\bibinfo {author} {\bibfnamefont {R.}~\bibnamefont
  {Skrotzki}}, \bibinfo {author} {\bibfnamefont {T.}~\bibnamefont
  {Herrmannsd{\"o}rfer}}, \bibinfo {author} {\bibfnamefont {V.}~\bibnamefont
  {Heera}}, \bibinfo {author} {\bibfnamefont {J.}~\bibnamefont {Fiedler}},
  \bibinfo {author} {\bibfnamefont {A.}~\bibnamefont {M{\"u}cklich}}, \bibinfo
  {author} {\bibfnamefont {M.}~\bibnamefont {Helm}}, \ and\ \bibinfo {author}
  {\bibfnamefont {J.}~\bibnamefont {Wosnitza}},\ }\href@noop {} {\bibfield
  {journal} {\bibinfo  {journal} {Low Temp. Phys.}\ }\textbf {\bibinfo {volume}
  {37}},\ \bibinfo {pages} {877} (\bibinfo {year} {2011})}\BibitemShut
  {NoStop}%
\bibitem [{\citenamefont {Prucnal}\ \emph {et~al.}(2019)\citenamefont
  {Prucnal}, \citenamefont {Heera}, \citenamefont {H{\"u}bner}, \citenamefont
  {Wang}, \citenamefont {Mazur}, \citenamefont {Grzybowski}, \citenamefont
  {Qin}, \citenamefont {Yuan}, \citenamefont {Voelskow}, \citenamefont
  {Skorupa}, \citenamefont {Rebohle}, \citenamefont {Helm}, \citenamefont
  {Sawicki},\ and\ \citenamefont {Zhou}}]{prucnal2019algaGe}%
  \BibitemOpen
  \bibfield  {author} {\bibinfo {author} {\bibfnamefont {S.}~\bibnamefont
  {Prucnal}}, \bibinfo {author} {\bibfnamefont {V.}~\bibnamefont {Heera}},
  \bibinfo {author} {\bibfnamefont {R.}~\bibnamefont {H{\"u}bner}}, \bibinfo
  {author} {\bibfnamefont {M.}~\bibnamefont {Wang}}, \bibinfo {author}
  {\bibfnamefont {G.~P.}\ \bibnamefont {Mazur}}, \bibinfo {author}
  {\bibfnamefont {M.~J.}\ \bibnamefont {Grzybowski}}, \bibinfo {author}
  {\bibfnamefont {X.}~\bibnamefont {Qin}}, \bibinfo {author} {\bibfnamefont
  {Y.}~\bibnamefont {Yuan}}, \bibinfo {author} {\bibfnamefont {M.}~\bibnamefont
  {Voelskow}}, \bibinfo {author} {\bibfnamefont {W.}~\bibnamefont {Skorupa}},
  \bibinfo {author} {\bibfnamefont {L.}~\bibnamefont {Rebohle}}, \bibinfo
  {author} {\bibfnamefont {M.}~\bibnamefont {Helm}}, \bibinfo {author}
  {\bibfnamefont {M.}~\bibnamefont {Sawicki}}, \ and\ \bibinfo {author}
  {\bibfnamefont {S.}~\bibnamefont {Zhou}},\ }\href@noop {} {\bibfield
  {journal} {\bibinfo  {journal} {Phys. Rev. Materials}\ }\textbf {\bibinfo
  {volume} {3}},\ \bibinfo {pages} {054802} (\bibinfo {year}
  {2019})}\BibitemShut {NoStop}%
\bibitem [{\citenamefont {Charnaya}\ \emph {et~al.}(1998)\citenamefont
  {Charnaya}, \citenamefont {Tien}, \citenamefont {Lin},\ and\ \citenamefont
  {Wur}}]{charnaya1998confinedga}%
  \BibitemOpen
  \bibfield  {author} {\bibinfo {author} {\bibfnamefont {E.~V.}\ \bibnamefont
  {Charnaya}}, \bibinfo {author} {\bibfnamefont {C.}~\bibnamefont {Tien}},
  \bibinfo {author} {\bibfnamefont {K.~J.}\ \bibnamefont {Lin}}, \ and\
  \bibinfo {author} {\bibfnamefont {C.~S.}\ \bibnamefont {Wur}},\ }\href@noop
  {} {\bibfield  {journal} {\bibinfo  {journal} {Phys. Rev. B}\ }\textbf
  {\bibinfo {volume} {58}},\ \bibinfo {pages} {467} (\bibinfo {year}
  {1998})}\BibitemShut {NoStop}%
\bibitem [{\citenamefont {Charnaya}\ \emph {et~al.}(2009)\citenamefont
  {Charnaya}, \citenamefont {Tien}, \citenamefont {Lee},\ and\ \citenamefont
  {Kumzerov}}]{charnaya2009gananoconfine}%
  \BibitemOpen
  \bibfield  {author} {\bibinfo {author} {\bibfnamefont {E.~V.}\ \bibnamefont
  {Charnaya}}, \bibinfo {author} {\bibfnamefont {C.}~\bibnamefont {Tien}},
  \bibinfo {author} {\bibfnamefont {M.~K.}\ \bibnamefont {Lee}}, \ and\
  \bibinfo {author} {\bibfnamefont {Y.~A.}\ \bibnamefont {Kumzerov}},\
  }\href@noop {} {\bibfield  {journal} {\bibinfo  {journal} {J. Phys.: Condens.
  Matter}\ }\textbf {\bibinfo {volume} {21}},\ \bibinfo {pages} {455304}
  (\bibinfo {year} {2009})}\BibitemShut {NoStop}%
\bibitem [{\citenamefont {Blase}\ \emph {et~al.}(2009)\citenamefont {Blase},
  \citenamefont {Bustarret}, \citenamefont {Chapelier}, \citenamefont {Klein},\
  and\ \citenamefont {Marcenat}}]{blase2009superIV}%
  \BibitemOpen
  \bibfield  {author} {\bibinfo {author} {\bibfnamefont {X.}~\bibnamefont
  {Blase}}, \bibinfo {author} {\bibfnamefont {E.}~\bibnamefont {Bustarret}},
  \bibinfo {author} {\bibfnamefont {C.}~\bibnamefont {Chapelier}}, \bibinfo
  {author} {\bibfnamefont {T.}~\bibnamefont {Klein}}, \ and\ \bibinfo {author}
  {\bibfnamefont {C.}~\bibnamefont {Marcenat}},\ }\href@noop {} {\bibfield
  {journal} {\bibinfo  {journal} {Nat. Mater.}\ }\textbf {\bibinfo {volume}
  {8}},\ \bibinfo {pages} {375} (\bibinfo {year} {2009})}\BibitemShut {NoStop}%
\bibitem [{\citenamefont {Carbotte}(1990)}]{carbotte1990bosonxchange}%
  \BibitemOpen
  \bibfield  {author} {\bibinfo {author} {\bibfnamefont {J.~P.}\ \bibnamefont
  {Carbotte}},\ }\href@noop {} {\bibfield  {journal} {\bibinfo  {journal} {Rev.
  Mod. Phys.}\ }\textbf {\bibinfo {volume} {62}},\ \bibinfo {pages} {1027}
  (\bibinfo {year} {1990})}\BibitemShut {NoStop}%
\bibitem [{\citenamefont {Tsu}\ \emph {et~al.}(1994)\citenamefont {Tsu},
  \citenamefont {Xiao}, \citenamefont {Kim}, \citenamefont {Hasa},
  \citenamefont {Birnbaum}, \citenamefont {Greene}, \citenamefont {Lin},\ and\
  \citenamefont {Chiang}}]{tsu1994siongembe}%
  \BibitemOpen
  \bibfield  {author} {\bibinfo {author} {\bibfnamefont {R.}~\bibnamefont
  {Tsu}}, \bibinfo {author} {\bibfnamefont {H.~Z.}\ \bibnamefont {Xiao}},
  \bibinfo {author} {\bibfnamefont {Y.-W.}\ \bibnamefont {Kim}}, \bibinfo
  {author} {\bibfnamefont {M.-A.}\ \bibnamefont {Hasa}}, \bibinfo {author}
  {\bibfnamefont {H.~K.}\ \bibnamefont {Birnbaum}}, \bibinfo {author}
  {\bibfnamefont {J.~E.}\ \bibnamefont {Greene}}, \bibinfo {author}
  {\bibfnamefont {D.-S.}\ \bibnamefont {Lin}}, \ and\ \bibinfo {author}
  {\bibfnamefont {T.-C.}\ \bibnamefont {Chiang}},\ }\href@noop {} {\bibfield
  {journal} {\bibinfo  {journal} {J. Appl. Phys.}\ }\textbf {\bibinfo {volume}
  {75}},\ \bibinfo {pages} {240} (\bibinfo {year} {1994})}\BibitemShut
  {NoStop}%
\bibitem [{\citenamefont {Kawabata}\ \emph {et~al.}(1989)\citenamefont
  {Kawabata}, \citenamefont {Ueba},\ and\ \citenamefont
  {Tatsuyama}}]{kawabata1989siheteroepi}%
  \BibitemOpen
  \bibfield  {author} {\bibinfo {author} {\bibfnamefont {H.}~\bibnamefont
  {Kawabata}}, \bibinfo {author} {\bibfnamefont {H.}~\bibnamefont {Ueba}}, \
  and\ \bibinfo {author} {\bibfnamefont {C.}~\bibnamefont {Tatsuyama}},\
  }\href@noop {} {\bibfield  {journal} {\bibinfo  {journal} {J. Appl. Phys.}\
  }\textbf {\bibinfo {volume} {66}},\ \bibinfo {pages} {634} (\bibinfo {year}
  {1989})}\BibitemShut {NoStop}%
\bibitem [{\citenamefont {Copel}\ \emph {et~al.}(1989)\citenamefont {Copel},
  \citenamefont {Reuter}, \citenamefont {Kaxiras},\ and\ \citenamefont
  {Tromp}}]{copel1989episurfactant}%
  \BibitemOpen
  \bibfield  {author} {\bibinfo {author} {\bibfnamefont {M.}~\bibnamefont
  {Copel}}, \bibinfo {author} {\bibfnamefont {M.~C.}\ \bibnamefont {Reuter}},
  \bibinfo {author} {\bibfnamefont {E.}~\bibnamefont {Kaxiras}}, \ and\
  \bibinfo {author} {\bibfnamefont {R.~M.}\ \bibnamefont {Tromp}},\ }\href@noop
  {} {\bibfield  {journal} {\bibinfo  {journal} {Phys. Rev. Lett.}\ }\textbf
  {\bibinfo {volume} {63}},\ \bibinfo {pages} {632} (\bibinfo {year}
  {1989})}\BibitemShut {NoStop}%
\bibitem [{\citenamefont {Eisenstein}(1954)}]{eisenstein1954gasupercond}%
  \BibitemOpen
  \bibfield  {author} {\bibinfo {author} {\bibfnamefont {J.}~\bibnamefont
  {Eisenstein}},\ }\href@noop {} {\bibfield  {journal} {\bibinfo  {journal}
  {Rev. Mod. Phys.}\ }\textbf {\bibinfo {volume} {26}},\ \bibinfo {pages} {277}
  (\bibinfo {year} {1954})}\BibitemShut {NoStop}%
\bibitem [{\citenamefont {Tedrow}\ and\ \citenamefont
  {Meservey}(1973)}]{Tedrow73}%
  \BibitemOpen
  \bibfield  {author} {\bibinfo {author} {\bibfnamefont {P.~M.}\ \bibnamefont
  {Tedrow}}\ and\ \bibinfo {author} {\bibfnamefont {R.}~\bibnamefont
  {Meservey}},\ }\href {\doibase 10.1103/PhysRevB.8.5098} {\bibfield  {journal}
  {\bibinfo  {journal} {Phys. Rev. B}\ }\textbf {\bibinfo {volume} {8}},\
  \bibinfo {pages} {5098} (\bibinfo {year} {1973})}\BibitemShut {NoStop}%
\bibitem [{\citenamefont {Moura}\ \emph {et~al.}(2017)\citenamefont {Moura},
  \citenamefont {Pirota}, \citenamefont {B\'{e}ron}, \citenamefont {Jesus},
  \citenamefont {Rosa}, \citenamefont {Tobia}, \citenamefont {Pagliuso},\ and\
  \citenamefont {de~Lima}}]{moura2017betaGa}%
  \BibitemOpen
  \bibfield  {author} {\bibinfo {author} {\bibfnamefont {K.~O.}\ \bibnamefont
  {Moura}}, \bibinfo {author} {\bibfnamefont {K.~R.}\ \bibnamefont {Pirota}},
  \bibinfo {author} {\bibfnamefont {F.}~\bibnamefont {B\'{e}ron}}, \bibinfo
  {author} {\bibfnamefont {C.~B.~R.}\ \bibnamefont {Jesus}}, \bibinfo {author}
  {\bibfnamefont {P.~F.~S.}\ \bibnamefont {Rosa}}, \bibinfo {author}
  {\bibfnamefont {D.}~\bibnamefont {Tobia}}, \bibinfo {author} {\bibfnamefont
  {P.~G.}\ \bibnamefont {Pagliuso}}, \ and\ \bibinfo {author} {\bibfnamefont
  {O.~F.}\ \bibnamefont {de~Lima}},\ }\href@noop {} {\bibfield  {journal}
  {\bibinfo  {journal} {Sci. Rep.}\ }\textbf {\bibinfo {volume} {7}},\ \bibinfo
  {pages} {15306} (\bibinfo {year} {2017})}\BibitemShut {NoStop}%
\bibitem [{\citenamefont {Teske}\ and\ \citenamefont
  {Drumheller}(1999)}]{teske1999gapolymorph}%
  \BibitemOpen
  \bibfield  {author} {\bibinfo {author} {\bibfnamefont {D.}~\bibnamefont
  {Teske}}\ and\ \bibinfo {author} {\bibfnamefont {J.~E.}\ \bibnamefont
  {Drumheller}},\ }\href@noop {} {\bibfield  {journal} {\bibinfo  {journal} {J.
  Phys.: Condens. Matter}\ }\textbf {\bibinfo {volume} {11}},\ \bibinfo {pages}
  {4935} (\bibinfo {year} {1999})}\BibitemShut {NoStop}%
\bibitem [{\citenamefont {Eley}\ \emph {et~al.}(2011)\citenamefont {Eley},
  \citenamefont {Gopalakrishnan}, \citenamefont {Goldbart},\ and\ \citenamefont
  {Mason}}]{eley2011nbisland}%
  \BibitemOpen
  \bibfield  {author} {\bibinfo {author} {\bibfnamefont {S.}~\bibnamefont
  {Eley}}, \bibinfo {author} {\bibfnamefont {S.}~\bibnamefont
  {Gopalakrishnan}}, \bibinfo {author} {\bibfnamefont {P.~M.}\ \bibnamefont
  {Goldbart}}, \ and\ \bibinfo {author} {\bibfnamefont {N.}~\bibnamefont
  {Mason}},\ }\href@noop {} {\bibfield  {journal} {\bibinfo  {journal} {Nat.
  Phys}\ }\textbf {\bibinfo {volume} {8}},\ \bibinfo {pages} {59} (\bibinfo
  {year} {2011})}\BibitemShut {NoStop}%
\bibitem [{\citenamefont {Cohen}(1964{\natexlab{a}})}]{cohen1964supersemi}%
  \BibitemOpen
  \bibfield  {author} {\bibinfo {author} {\bibfnamefont {M.~L.}\ \bibnamefont
  {Cohen}},\ }\href@noop {} {\bibfield  {journal} {\bibinfo  {journal} {Phys.
  Rev.}\ }\textbf {\bibinfo {volume} {134}},\ \bibinfo {pages} {A511} (\bibinfo
  {year} {1964}{\natexlab{a}})}\BibitemShut {NoStop}%
\bibitem [{\citenamefont {Cohen}(1964{\natexlab{b}})}]{cohen1964supersemi2}%
  \BibitemOpen
  \bibfield  {author} {\bibinfo {author} {\bibfnamefont {M.~L.}\ \bibnamefont
  {Cohen}},\ }\href@noop {} {\bibfield  {journal} {\bibinfo  {journal} {Rev.
  Mod. Phys.}\ }\textbf {\bibinfo {volume} {36}},\ \bibinfo {pages} {240}
  (\bibinfo {year} {1964}{\natexlab{b}})}\BibitemShut {NoStop}%
\end{thebibliography}%

\end{document}